\documentclass[twocolumn,showpacs,preprintnumbers,amsmath,amssymb]{revtex4}
\begin{document}
\draft{}
\bibliographystyle{try}

\topmargin 0.1cm

 \newcounter{univ_counter}
 \setcounter{univ_counter} {0}

\addtocounter{univ_counter} {1} 
\edef\INFNGE{$^{\arabic{univ_counter}}$ } 

\addtocounter{univ_counter} {1} 
\edef\JLAB{$^{\arabic{univ_counter}}$ } 

\addtocounter{univ_counter} {1} 
\edef\MSU{$^{\arabic{univ_counter}}$ } 

\addtocounter{univ_counter} {1} 
\edef\ASU{$^{\arabic{univ_counter}}$ } 

\addtocounter{univ_counter} {1} 
\edef\UCLA{$^{\arabic{univ_counter}}$ } 

\addtocounter{univ_counter} {1} 
\edef\CMU{$^{\arabic{univ_counter}}$ } 

\addtocounter{univ_counter} {1} 
\edef\CUA{$^{\arabic{univ_counter}}$ } 

\addtocounter{univ_counter} {1} 
\edef\SACLAY{$^{\arabic{univ_counter}}$ } 

\addtocounter{univ_counter} {1} 
\edef\CNU{$^{\arabic{univ_counter}}$ } 

\addtocounter{univ_counter} {1} 
\edef\UCONN{$^{\arabic{univ_counter}}$ } 

\addtocounter{univ_counter} {1} 
\edef\DUKE{$^{\arabic{univ_counter}}$ } 

\addtocounter{univ_counter} {1} 
\edef\EDINBURGH{$^{\arabic{univ_counter}}$ } 

\addtocounter{univ_counter} {1} 
\edef\FIU{$^{\arabic{univ_counter}}$ } 

\addtocounter{univ_counter} {1} 
\edef\FSU{$^{\arabic{univ_counter}}$ } 

\addtocounter{univ_counter} {1} 
\edef\GWU{$^{\arabic{univ_counter}}$ } 

\addtocounter{univ_counter} {1} 
\edef\GLASGOW{$^{\arabic{univ_counter}}$ } 

\addtocounter{univ_counter} {1} 
\edef\INFNFR{$^{\arabic{univ_counter}}$ } 

\addtocounter{univ_counter} {1} 
\edef\ORSAY{$^{\arabic{univ_counter}}$ } 

\addtocounter{univ_counter} {1} 
\edef\ITEP{$^{\arabic{univ_counter}}$ } 

\addtocounter{univ_counter} {1} 
\edef\JMU{$^{\arabic{univ_counter}}$ } 

\addtocounter{univ_counter} {1} 
\edef\KYUNGPOOK{$^{\arabic{univ_counter}}$ } 

\addtocounter{univ_counter} {1} 
\edef\MIT{$^{\arabic{univ_counter}}$ } 

\addtocounter{univ_counter} {1} 
\edef\UMASS{$^{\arabic{univ_counter}}$ } 

\addtocounter{univ_counter} {1} 
\edef\UNH{$^{\arabic{univ_counter}}$ } 

\addtocounter{univ_counter} {1} 
\edef\NSU{$^{\arabic{univ_counter}}$ } 

\addtocounter{univ_counter} {1} 
\edef\OHIOU{$^{\arabic{univ_counter}}$ } 

\addtocounter{univ_counter} {1} 
\edef\ODU{$^{\arabic{univ_counter}}$ } 

\addtocounter{univ_counter} {1} 
\edef\PITT{$^{\arabic{univ_counter}}$ } 

\addtocounter{univ_counter} {1} 
\edef\ROMA{$^{\arabic{univ_counter}}$ } 

\addtocounter{univ_counter} {1} 
\edef\RPI{$^{\arabic{univ_counter}}$ } 

\addtocounter{univ_counter} {1} 
\edef\RICE{$^{\arabic{univ_counter}}$ } 

\addtocounter{univ_counter} {1} 
\edef\URICH{$^{\arabic{univ_counter}}$ } 

\addtocounter{univ_counter} {1} 
\edef\SCAROLINA{$^{\arabic{univ_counter}}$ } 

\addtocounter{univ_counter} {1} 
\edef\UTEP{$^{\arabic{univ_counter}}$ } 

\addtocounter{univ_counter} {1} 
\edef\VT{$^{\arabic{univ_counter}}$ } 

\addtocounter{univ_counter} {1} 
\edef\VIRGINIA{$^{\arabic{univ_counter}}$ } 

\addtocounter{univ_counter} {1} 
\edef\WM{$^{\arabic{univ_counter}}$ } 

\addtocounter{univ_counter} {1} 
\edef\YEREVAN{$^{\arabic{univ_counter}}$ } 

% \title{{\large Baryon resonance analysis from two-pion electroproduction}}
\title{{\large Measurement of $ep \rightarrow e^{'}p\pi^{+}\pi^{-}$: 
experimental procedures and baryon resonance analysis}}
%%%%%%%%%%%%%%%%%%%% authors %%%%%%%%% 
 \author{ 
M.~Ripani,\INFNGE\
V.D.~Burkert,\JLAB\
V.~Mokeev,\MSU\
M.~Battaglieri,\INFNGE\
R.~De~Vita,\INFNGE\
E.~Golovach,\MSU\
M.~Taiuti,\INFNGE\
G.~Adams,\RPI\
E.~Anciant,\SACLAY\
M.~Anghinolfi,\INFNGE
% ...
} 

\address{\INFNGE INFN, Sezione di Genova, 16146 Genova, Italy}
\address{\JLAB Thomas Jefferson National Accelerator Facility, Newport News, Virginia 23606}
\address{\MSU Moscow State University, 119899 Moscow, Russia}
\address{\ASU Arizona State University, Tempe, Arizona 85287}
\address{\UCLA University of California at Los Angeles, Los Angeles, California  90095}
\address{\CMU Carnegie Mellon University, Pittsburgh, Pennsylvania 15213}
\address{\CUA Catholic University of America, Washington, D.C. 20064}
\address{\SACLAY CEA-Saclay, Service de Physique Nucl\'eaire, F91191 Gif-sur-Yvette, Cedex, France}
\address{\CNU Christopher Newport University, Newport News, Virginia 23606}
\address{\UCONN University of Connecticut, Storrs, Connecticut 06269}
\address{\DUKE Duke University, Durham, North Carolina 27708}
\address{\EDINBURGH Edinburgh University, Edinburgh EH9 3JZ, United Kingdom}
\address{\FIU Florida International University, Miami, Florida 33199}
\address{\FSU Florida State University, Tallahasee, Florida 32306}
\address{\GWU The George Washington University, Washington, DC 20052}
\address{\GLASGOW University of Glasgow, Glasgow G12 8QQ, United Kingdom}
\address{\INFNFR INFN, Laboratori Nazionali di Frascati, PO 13, 00044 Frascati, Italy}
\address{\ORSAY Institut de Physique Nucleaire ORSAY, IN2P3 BP 1, 91406 Orsay, France}
\address{\ITEP Institute of Theoretical and Experimental Physics, Moscow, 117259, Russia}
\address{\JMU James Madison University, Harrisonburg, Virginia 22807}
\address{\KYUNGPOOK Kyungpook National University, Daegu 702-701, South Korea}
\address{\MIT Massachusetts Institute of Technology, Cambridge, Massachusetts  02139}
\address{\UMASS University of Massachusetts, Amherst, Massachusetts  01003}
\address{\UNH University of New Hampshire, Durham, New Hampshire 03824}
\address{\NSU Norfolk State University, Norfolk, Virginia 23504}
\address{\OHIOU Ohio University, Athens, Ohio 45701}
\address{\ODU Old Dominion University, Norfolk, Virginia 23529}
\address{\PITT University of Pittsburgh, Pittsburgh, Pennsylvania 15260}
\address{\ROMA Universita' di ROMA III, 00146 Roma, Italy}
\address{\RPI Rensselaer Polytechnic Institute, Troy, New York 12180}
\address{\RICE Rice University, Houston, Texas 77005}
\address{\URICH University of Richmond, Richmond, Virginia 23173}
\address{\SCAROLINA University of South Carolina, Columbia, South Carolina 29208}
\address{\UTEP University of Texas at El Paso, El Paso, Texas 79968}
\address{\VT Virginia Polytechnic Institute and State University, Blacksburg, Virginia   24061}
\address{\VIRGINIA University of Virginia, Charlottesville, Virginia 22901}
\address{\WM College of William and Mary, Williamsburg, Virginia 23187}
\address{\YEREVAN Yerevan Physics Institute, 375036 Yerevan, Armenia}
% 
 
%The Southeastern Universities Research Association (SURA) operates the 
%Thomas Jefferson National Accelerator Facility for the United States 
%Department of Energy under contract DE-AC05-84ER40150. 
% 
% 

\date{\today}

\begin{abstract}
The cross section for the reaction $ e p \rightarrow e^{\prime} p \pi^{+} \pi^{-}$ 
was measured in the resonance region for 1.4$<$W$<$2.1 GeV and
0.5$<Q^{2}<$1.5 GeV$^{2}$/c$^{2}$ using the CLAS detector at Jefferson Laboratory.
The data show resonant 
structures not visible in previous experiments.
The comparison of our 
data to a phenomenological prediction using available information
on $N^{*}$ and $\Delta$ states shows an evident discrepancy.
A better description of the data is obtained either by a sizeable change of the
properties of the $P_{13}$(1720) resonance or by introducing a new
baryon state, not reported in published analyses. 
\end{abstract}

\pacs{13.60.Le, 13.40.Gp, 14.20.Gk}

\maketitle

Electromagnetic excitation of nucleon resonances is sensitive to the 
spin and spatial structure of the transition, which in turn is
connected to fundamental properties of baryon structure,
like spin-flavor symmetries, confinement, and 
effective degrees of freedom.
In the mass region above 1.6 GeV, many overlapping baryon states
are present, and some of them are not well known;  
measurement of the transition form factors of these states 
is important for our understanding of the internal dynamics
of baryons.
Many of these high-mass excited states 
tend to decouple from the single-meson 
channels and to decay predominantly 
into multi-pion channels, such as $\Delta \pi$ or $N \rho$, leading 
to $N \pi \pi$  final states \cite{Pdg96}. 
Moreover, quark models
with approximate (or ``broken'') SU(6)$\otimes$O(3) symmetry \cite{Kon80,Gia90} 
predict more states than have been found experimentally; 
QCD mixing effects could decouple these unobserved states 
from the pion-nucleon channel \cite{Kon80} while
strongly coupling them to two-pion channels \cite{Kon80,Cap94,Sta93}. 
These states would therefore not be observable in reactions with
$\pi N$ in the initial or final state.
Other models, with different symmetry properties and a reduced
number of degrees of freedom, as e.g. in ref. \cite{Kir97}, 
predict fewer states. 
Experimental searches for at least some of the
   ``missing'' states predicted by the symmetric quark models, which are
   not predicted by models using alternative symmetries, are crucial in
   discriminating between these models.
Electromagnetic amplitudes for some missing states are predicted
to be sizeable \cite{Kon80} as well.
Therefore, exclusive double-pion electroproduction 
is a fundamental tool in measuring poorly known states and 
possibly observing new ones. 

In this paper we report a measurement 
of the $e p \rightarrow e^{\prime} p \pi^{+} \pi^{-}$ reaction 
studied with the CEBAF Large Acceptance Spectrometer (CLAS) 
at Jefferson Lab. 
CLAS consists of a six-coil superconducting
    magnet producing an approximately toroidal magnetic field,
    allowing detection of 
   electrons and hadrons and full 4-momentum 
   recontruction. 
   Three sets of drift chambers allow the determination
   of the momenta of the charged particles with polar angles 
   from 10$^o$ to 140$^o$. A complete coverage of scintillators
   allows the discrimination of particles by a time-of-flight 
   technique described in ref.~\cite{Smi99}.   
   The magnetic field was set to bent positive particles outwards,
   away from the primary beam. 
We analyzed data taken in the so-called
'e1c' running period, corresponding to about two months 
of data taking in the spring of 1999.
Beam currents of 
a few nA were delivered to Hall B on a liquid-hydrogen target, 
corresponding to luminosities up to $4 \times 10^{33}$~cm$^{-2}$s$^{-1}$. 
The beam energies selected for this analysis 
were: 2.567 GeV at a torus current of
1500 A ('low' field), with the goal of obtaining
two $Q^{2}$ bins, between 0.5 and 0.8
$GeV^{2}/c^{2}$ and between 0.8 and 1.1
$GeV^{2}/c^{2}$, with $W$ up to 1.9 GeV;
4.247 GeV at a field of 2250 A ('low' field),
to obtain $Q^{2}$ between 1.1 and 1.5 $GeV^{2}/c^{2}$ ,
with  $W$ up to 2.1 GeV.
Important features
of the CLAS~\cite{Bro99} are its large kinematic coverage for 
multi-charged-particle final states
and its good momentum resolution ($\Delta p/p \sim$1\%). Using an
inclusive electron trigger,
many exclusive hadronic final states
were measured simultaneously. Scattered electrons were identified through
cuts on the calorimeter energy loss and the Cerenkov photo-electron distribution.
Different channels were separated through
particle identification using time-of-flight information and other 
kinematic cuts.

The hardware trigger in CLAS was based on a coincidence
between Cherenkov counter and electromagnetic forward
calorimeter. The discriminator threshold in the Cherenkov
was put at a signal level of less than a single photoelectron 
produced on the photocathode, to avoid losing good events
while rejecting most of the pion background. The energy 
threshold in the
calorimeter was set to cut off events as close as
possible to the kinematic edges of the W and Q$^{2}$
domain covered by the measurement, but 
sufficiently high to keep the low energy contamination 
        due to hadronically interacting particles at a reasonable 
       limit.
A cut on the energy deposited in the calorimeter
was applied to eliminate part of the remaining
pion background, mostly due to knock-off electrons
giving a signal in the Cherenkov counter.
We studied the effect of this cut on
the electron detection efficiency, deriving a
correction that was applied to the data.
The Cherenkov efficiency was also studied by means
of the shape of the photoelectron distribution in 
each module. This way, we evaluated the effect of
a cut on the photoelectron number on the electron
detection efficiency on each single Cherenkov module,
as well as the module-to-module uniformity, deriving
the appropriate corrections to be applied to the data.

\begin{figure}[h]
\vspace{5cm}                                                                  
\includegraphics{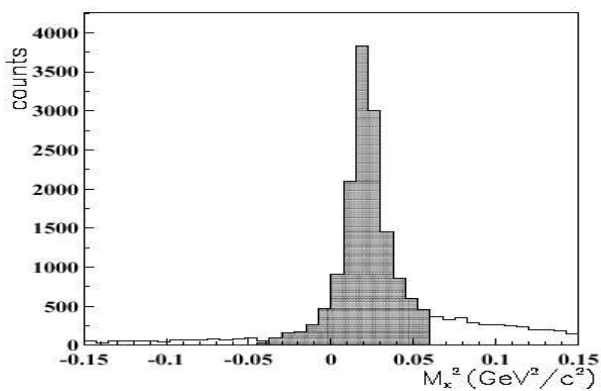}
\caption[]{Missing mass for detection of $e p \pi^{+}$, at
$W$=1.6-1.7 GeV and at $Q^{2}$=0.5-0.8~(GeV/c)$^2$. 
The tail at higher mass is due to radiative effects and to multiple
pion production.
The hatched area represents the adopted cut.}
\label{fig:MissMass}
\end{figure}

Another cut was applied in the $\beta$ versus momentum plot,
to select pions and protons in the reaction analyzed.
All Time of Flight (ToF) scintillator paddles were carefully inspected to make
sure that the cut applied was appropriate for all of them.
To identify the $e^{\prime} p \pi^{+} \pi^{-}$ final state,
we used the missing-mass technique, 
requiring detection in CLAS of at least $e p \pi^{+}$.
The good resolution allowed selection of the exclusive
final state, $ep \pi^{+} \pi^{-}$.  After applying all cuts, our data sample
included about $2 \times 10^{5}$ two-pion events. 
Fig.~\ref{fig:MissMass} shows the missing mass distribution
obtained from CLAS, together with the applied cut. 
The tail at higher mass is due to radiative effects and to multiple
pion production.

In order to check the stability of CLAS in the detection
of different reactions, we defined a set of histograms
representing various reaction yields, i.e. electron inclusive,
electron-proton inclusive, electron elastic (W cut), 
electron-proton elastic (W plus $\theta$-$\phi$ correlation
cut), electron-proton from the $\Delta$ (W cut) and finally
electron-proton-$\pi^{+}$ with a cut on the missing $\pi^{-}$, 
each single yield being normalised to the Faraday cup charge
obtained from the so called ``live-gated'' signal, where the
signal from the Faraday cup is only integrated during the
live-time: this way, the charge is already corrected for
the data acquisition dead-time. Such normalised yields
were calculated, using the PID procedures described in the
previous subsection, for each data file inside a run
(a run being typically a data taking over a period of an hour), 
therefore
providing a very accurate monitoring of stability even
inside a single run. ``Good files'' were selected requiring 
that the normalised yield for a single file should not deviate
more than a few percent from the average. Subsequent analysis
was performed only on the good files. 

To obtain the cross section from the
raw data, it is necessary to correct for
detector non-uniformity, which has origin
both in the geometry and in the response of the equipment
to different particles.
The geometrical and
kinematic non-uniformity can be very well described by
means of fiducial cuts that describe regions of the detector
where the response of the various subsystems is well
known. Fiducial cuts will eliminate dead regions
like the torus coils in the first place. The detector
response inside the fiducial regions was simulated using a 
      GEANT-based representation of the detector containing a 
      detailed description of particle interaction with the various
      subsystems.
To evaluate all detector corrections, we divided the particle yield
into kinematic bins defined through a complete set of independent 
kinematic variables of the hadronic state, plus $W$ and $Q^2$. 
This corresponds to binning all four-momenta of the particles involved
(with the exception of the electron, for which no binning
is applied to $\phi$ azimuthal angle for symmetry reasons).
This way, the folding of the detector response 
to the cross sections is done in narrow kinematic regions, 
where the cross section variation will be limited. The dependence of
the results on the assumed particle distributions
will therefore be much reduced.
However, events still have to be generated according
to a realistic Monte Carlo, in order to minimize
the model dependence of acceptance and efficiency. 

\begin{figure}[h]
\vspace{6cm}                                                                  
\includegraphics{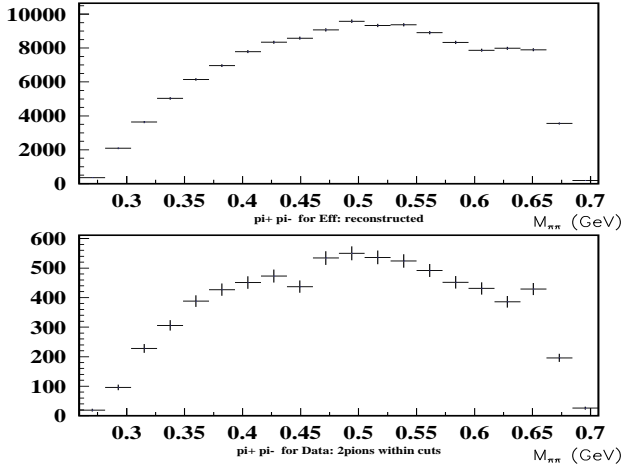}
\caption{Simulated (top) and measured (bottom) 
invariant mass for the pion-pion pair, when detecting $e p \pi^{+}$,
at 2.567 GeV beam energy, 1500 A field, W between 1.6 and 1.625 GeV and $Q^{2}$ between
0.5 and 0.8 GeV$^{2}$/c$^{2}$.
\label{fig:data_MC_1}}
\end{figure}

\begin{figure}[h]
\vspace{6cm}                                                                  
\includegraphics{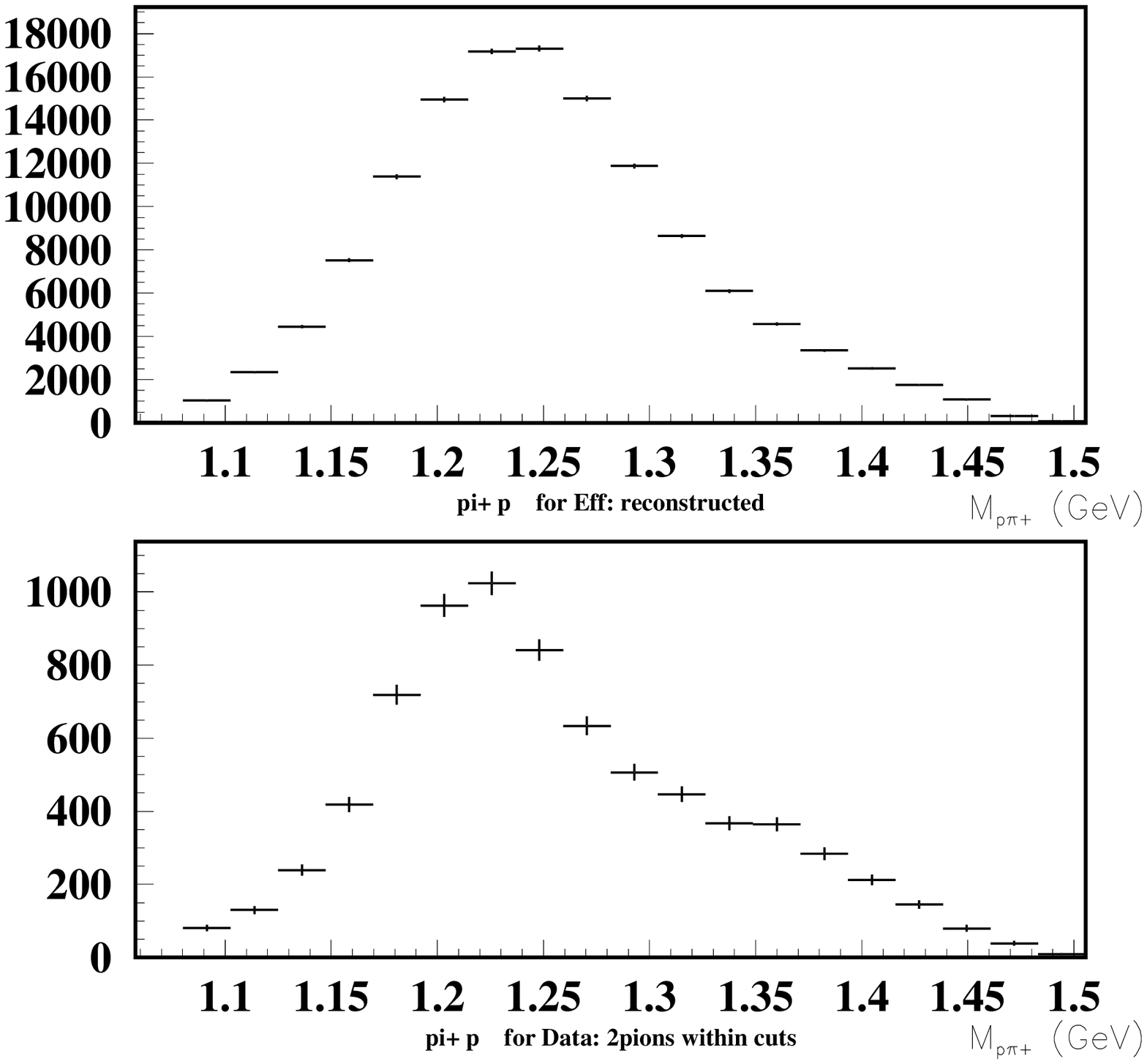}
\caption{Simulated (top) and measured (bottom) 
invariant mass for the proton-$\pi^{+}$ pair, when detecting $e p \pi^{+}$,
at 2.567 GeV beam energy, 1500 A field, W between 1.6 and 1.625 GeV and $Q^{2}$ between
0.5 and 0.8 GeV$^{2}$/c$^{2}$.
\label{fig:data_MC_2}}
\end{figure}

\begin{figure}[h]
\vspace{6cm}                                                                  
\includegraphics{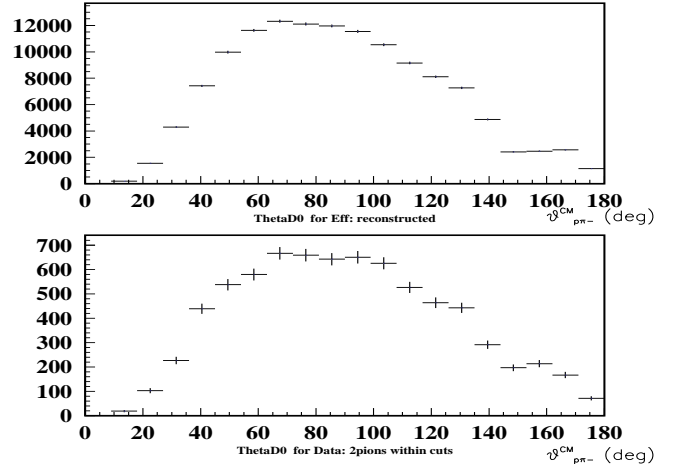}
\caption{Simulated (top) and measured (bottom) 
angular distribution of the $p-\pi^{-}$ pair in the CM system, when detecting $e p \pi^{+}$,
at 2.567 GeV beam energy, 1500 A field, W between 1.6 and 1.625 GeV and $Q^{2}$ between
0.5 and 0.8 GeV$^{2}$/c$^{2}$.
\label{fig:data_MC_3}}
\end{figure}

The event generator used for simulations in this
experiment contains several electroproduction cross
sections, including single, double and triple pion
electroproduction. The code relies on cross section
tables that describe measured total and differential
cross sections from the literature, scaled by a virtual
photon flux and a dipole form factor to provide a
reasonable fall off with $Q^{2}$.
Therefore our code gives a realistic description
of cross sections and their relative weights,
as well as backgrounds generated from competing
channels. Radiative effects were also included
in 
         the simulation. As an example of how the Monte Carlo 
         reproduces the main
features seen in the data, we reproduce here the comparison
of some kinematic distributions from the data 
at 2.567 GeV beam energy, 1500 A field,
with the corresponding simulation. As specific example
we report the case of the $W$ bin 1.6-1.625 GeV, and $Q^{2}$ between
0.5 and 0.8 GeV$^{2}$/c$^{2}$. Figure~\ref{fig:data_MC_1} shows
the invariant mass distribution for the pion-pion pair.
Figure~\ref{fig:data_MC_2} shows
the invariant mass distribution for the proton-$\pi^{+}$ pair.
Figure~\ref{fig:data_MC_3} shows
the CM angle of the $p-\pi^{-}$ pair (which would correspond to the
CM angle of a $\Delta^{0}$ in the specific case that a $\Delta^{0}$
is produced).

A particularly important issue regards the percentage
of events lost in the binning process due to bins with
zero acceptance: of course, for such bins it is not
possible to correct the data and obtain a cross section;
being the kinematic variable space multidimensional in
the case of double pion production, extrapolating the
cross sections from neighboring bins with non-zero acceptance to bins
with zero acceptance can be unreliable.
Therefore, we checked carefully, using our realistic
Monte Carlo, the percentage of cross section lost because
of bins with zero acceptance. It turned out that with
the adopted binning, the percentage of unmeasured cross
section in bins with vanishing acceptance was below 10 \%
and typically of the order of a few percent. Actually, the
fact that a cell has zero acceptance or efficiency may
be just connected to insufficient statistics in the
simulation, as many cells end up with very few generated
events; nevertheless, for those cells it is not possible
to perform a correction to the data.
To get the final cross sections, we therefore performed an
extrapolation to the empty bins using the MonteCarlo distributions
as an estimate of the missing cross section. This extrapolation
is typically only a few percent and we quoted as systematic
error a quantity equal to one half of the extrapolation,
assuming that the extrapolated cross section may be
wrong by plus or minus 50 \%, due to the assumptions
in the MonteCarlo.

The range of invariant hadronic center-of-mass (CM) 
energy $W$ (in 25 MeV bins) was
1.4-1.9~GeV for the first two bins in the invariant momentum
transfer $Q^{2}$, 0.5-0.8~(GeV/c)$^2$ and 0.8-1.1~(GeV/c)$^2$, and 
1.4-2.1~GeV for the highest $Q^{2}$ bin, 1.1-1.5~(GeV/c)$^2$. 
Following the procedures schematically described above, 
data were corrected for acceptance, reconstruction efficiency, 
radiative effects, and 
empty target counts. They were further binned in the following 
set of hadronic CM variables: 
invariant mass of the $p \pi^{+}$ pair (10 bins), invariant mass 
of the $\pi^{+} \pi^{-}$ pair (10 bins), 
$\pi^{-}$ polar angle $\theta$ (10 bins), azimuthal angle $\phi$ (5 bins), 
and rotation freedom $\psi$ of the $p \pi^{+}$ pair with respect to
the hadronic plane (5 bins).
The fully differential cross section is
of the form: 
\begin{eqnarray} 
\frac{d\sigma}{dWdQ^{2}dM_{p \pi^{+}}dM_{\pi^{+}\pi^{-}}d\cos\theta_{\pi^{-}}d\phi_{\pi^{-}}d\psi_{p \pi^{+}}} = \nonumber \\
\Gamma_{v}\frac{d\sigma_{v}}{dM_{p \pi^{+}}dM_{\pi^{+}\pi^{-}}d\cos\theta_{\pi^{-}}d\phi_{\pi^{-}}d\psi_{p \pi^{+}}} =
\Gamma_{v}\frac{d\sigma_{v}}{d\tau}
\end{eqnarray} 
\begin{equation} 
\Gamma_{v} =
\frac{\alpha}{4 \pi}\frac{1}{E^{2}M_{p}^{2}}
\frac{W(W^{2}-M_{p}^{2})}{(1-\epsilon)Q^{2}}
\end{equation}
\noindent where $\Gamma_{v}$ is the virtual photon flux,
$\frac{d\sigma_{v}}{d\tau}$ is the virtual photon cross section,
$\alpha$ is the fine structure constant,
$E$ is the electron beam energy,
$M_{p}$ is the proton mass, and $\epsilon$ is the virtual photon
transverse polarization \cite{Ama89}.
\begin{figure}[h]
\vspace{6.0cm}                                                                  
\includegraphics{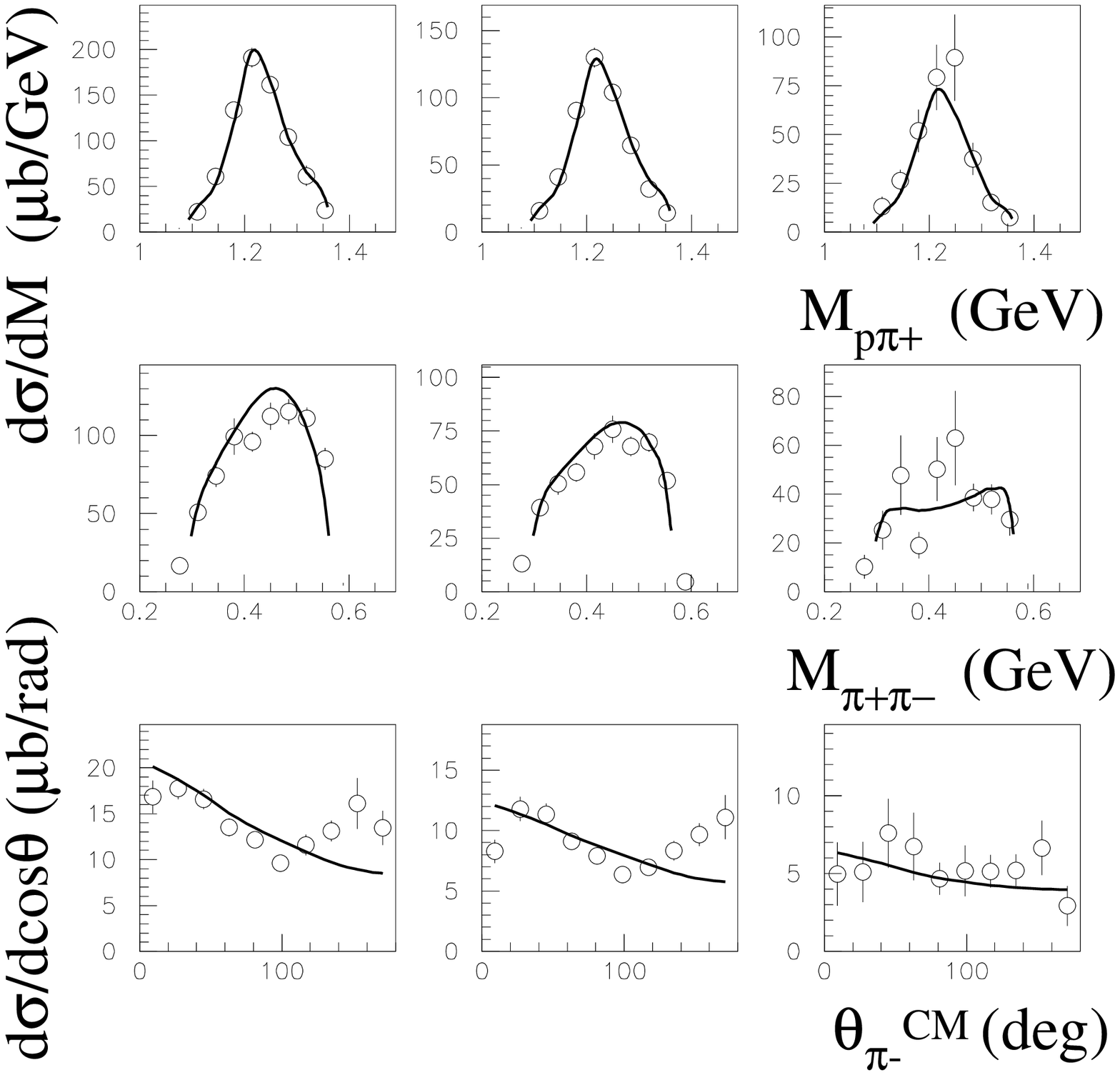}
%blinp13065-130_171_P11.ps hscale=48 vscale=35 hoffset=-12 voffset=-60}
\caption[]{$\frac{d\sigma_{v}}{dM_{p \pi^{+}}}$, 
$\frac{d\sigma_{v}}{dM_{\pi^{+}\pi^{-}}}$, 
and $\frac{d\sigma_{v}}{dcos\theta_{\pi^{-}}}$
from CLAS (from top to bottom) at $W$=1.5-1.525~GeV
and for the three mentioned $Q^{2}$ intervals (left to right).
The error bars include statistical errors only. 
The curves represent our step (A) reference calculations.
}
\label{fig:sqtm1}
\end{figure}
\begin{figure}[h]
\vspace{6.0cm}                                                                  
\includegraphics{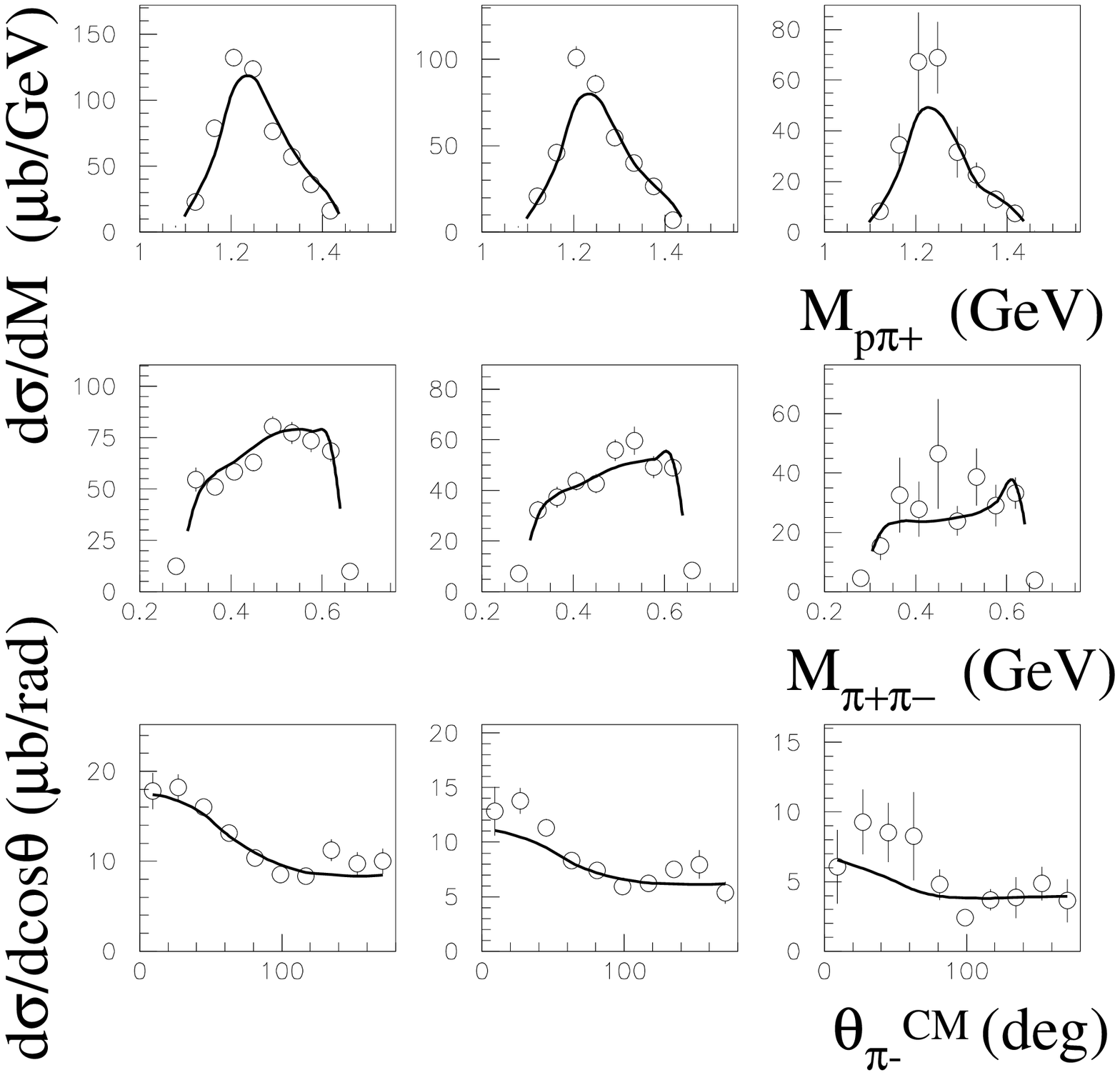}
%blinp13065-130_171_P11.ps hscale=48 vscale=35 hoffset=-12 voffset=-60}
\caption[]{$\frac{d\sigma_{v}}{dM_{p \pi^{+}}}$, 
$\frac{d\sigma_{v}}{dM_{\pi^{+}\pi^{-}}}$, 
and $\frac{d\sigma_{v}}{dcos\theta_{\pi^{-}}}$
from CLAS (from top to bottom) at $W$=1.575-1.6~GeV
and for the three mentioned $Q^{2}$ intervals (left to right).
The error bars include statistical errors only. 
The curves represent our step (A) reference calculations.
}
\label{fig:sqtm2}
\end{figure}
\begin{figure}[h]
\vspace{6.0cm}                                                                  
\includegraphics{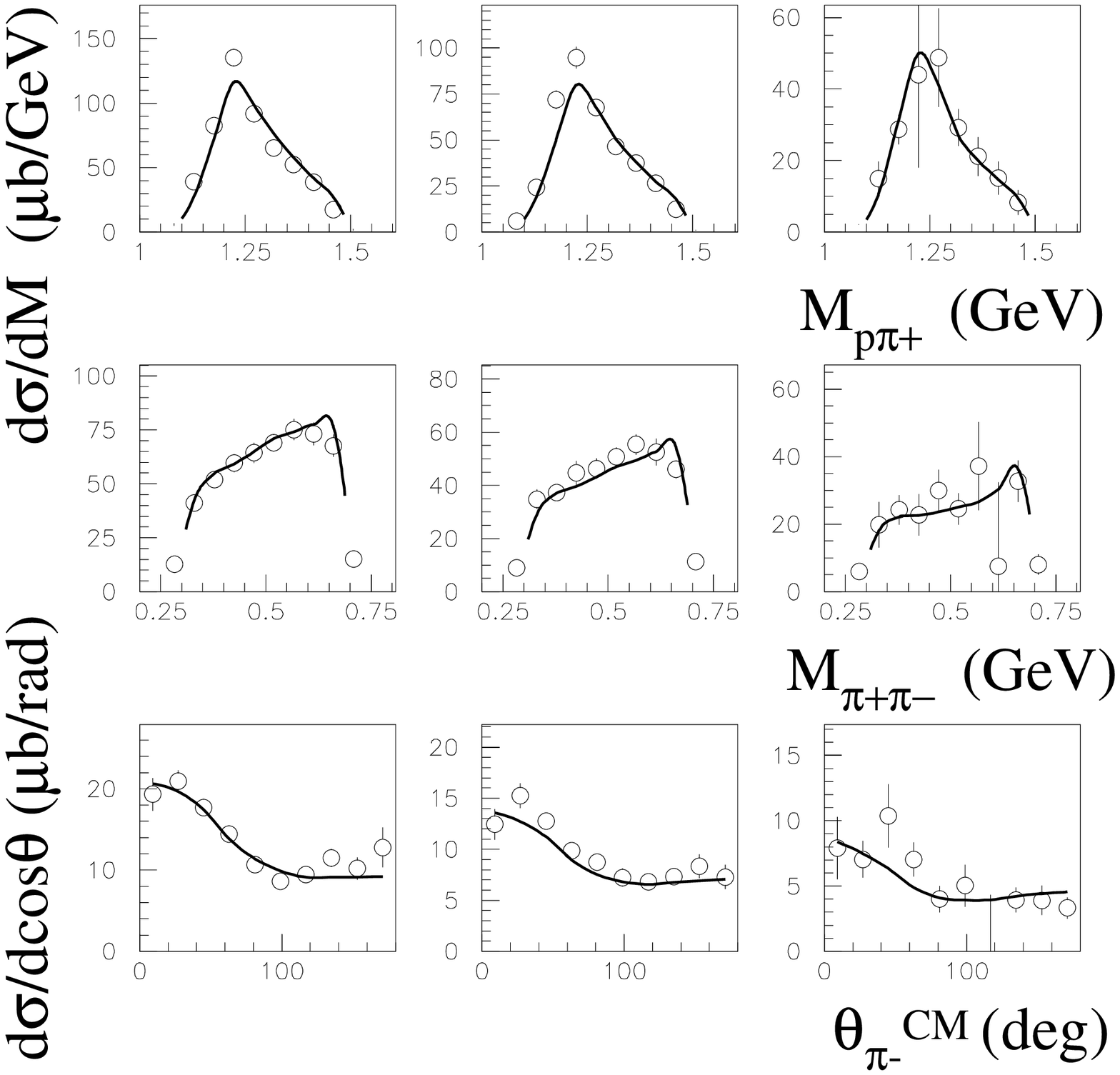}
%blinp13065-130_171_P11.ps hscale=48 vscale=35 hoffset=-12 voffset=-60}
\caption[]{$\frac{d\sigma_{v}}{dM_{p \pi^{+}}}$, 
$\frac{d\sigma_{v}}{dM_{\pi^{+}\pi^{-}}}$, 
and $\frac{d\sigma_{v}}{dcos\theta_{\pi^{-}}}$
from CLAS (from top to bottom) at $W$=1.625-1.65~GeV
and for the three mentioned $Q^{2}$ intervals (left to right).
The error bars include statistical errors only. 
The curves represent our step (A) reference calculations.
}
\label{fig:sqtm3}
\end{figure}
\begin{figure}[h]
\vspace{6.0cm}                                                                  
\includegraphics{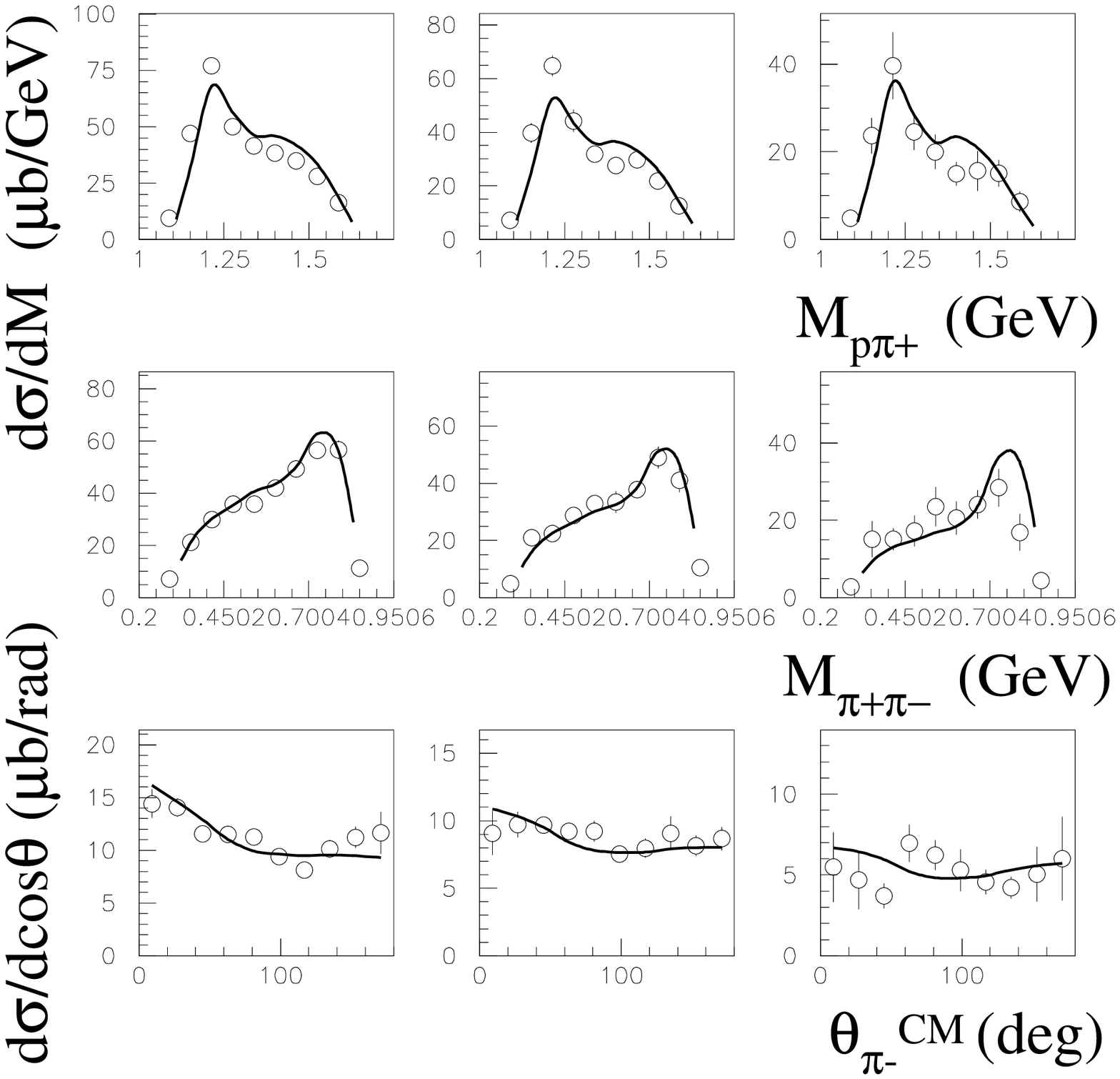}
%blinp13065-130_171_P11.ps hscale=48 vscale=35 hoffset=-12 voffset=-60}
\caption[]{$\frac{d\sigma_{v}}{dM_{p \pi^{+}}}$, 
$\frac{d\sigma_{v}}{dM_{\pi^{+}\pi^{-}}}$, 
and $\frac{d\sigma_{v}}{dcos\theta_{\pi^{-}}}$
from CLAS (from top to bottom) at $W$=1.775-1.8~GeV
and for the three mentioned $Q^{2}$ intervals (left to right).
The error bars include statistical errors only. 
The curves represent our step (A) reference calculations.
}
\label{fig:sqtm4}
\end{figure}
\begin{figure}[h]
\vspace{6.0cm}                                                                  
\includegraphics{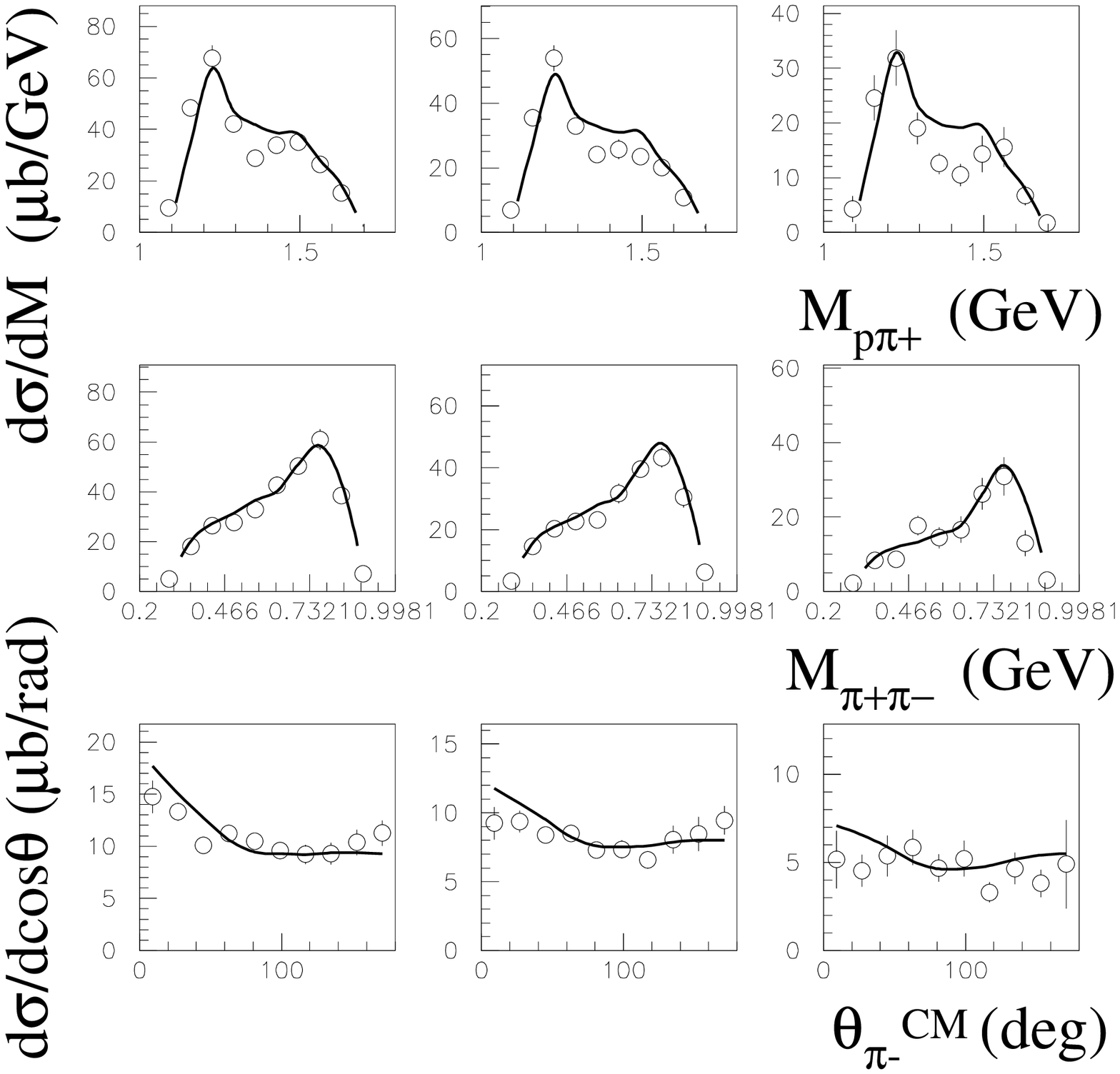}
%blinp13065-130_171_P11.ps hscale=48 vscale=35 hoffset=-12 voffset=-60}
\caption[]{$\frac{d\sigma_{v}}{dM_{p \pi^{+}}}$, 
$\frac{d\sigma_{v}}{dM_{\pi^{+}\pi^{-}}}$, 
and $\frac{d\sigma_{v}}{dcos\theta_{\pi^{-}}}$
from CLAS (from top to bottom) at $W$=1.825-1.85~GeV
and for the three mentioned $Q^{2}$ intervals (left to right).
The error bars include statistical errors only. 
The curves represent our step (A) reference calculations.
}
\label{fig:sqtm5}
\end{figure}
%
%Systematic uncertainties were estimated as a function of $W$ and $Q^{2}$.
%The main sources were acceptance modeling, finite integration steps,
%and modeling of the radiative corrections, each one being at the 3 to 10\% level. 
%Each of the various cuts applied (fiducial, missing mass, etc.) contributed 2 to 5\%.
%In Fig.~\ref{fig:Xsec_allq2_data_and_nominal_and_pipi} (left)
%we report the total virtual photon cross section as a function of
%$W$ for all $Q^{2}$ intervals analyzed. 
%The CLAS data points clearly exhibit structures, 
%not visible in previous data \cite{Eck73}
%due to limited statistical accuracy.
%
\begin{figure}[h]
\vspace{5.cm}                                                                  
\includegraphics{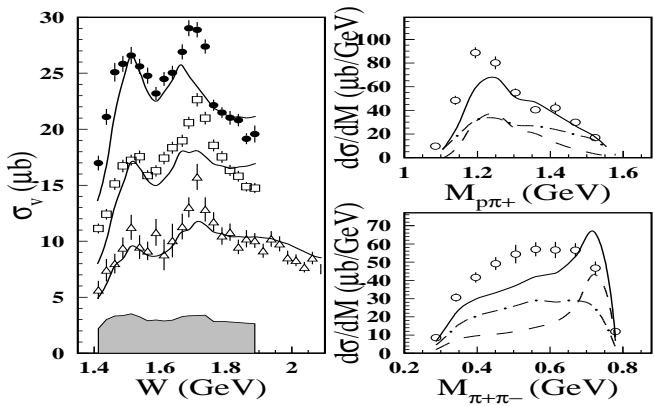}
%w_sqtm_and_mpipi_carman.ps hscale=48 vscale=30 hoffset=-12 voffset=-50}
\caption[]{Left: Total cross section for 
$\gamma_{v} p \rightarrow p \pi^{+} \pi^{-}$ as a function of $W$.
Data from CLAS are shown at $Q^{2}$=0.5-0.8~(GeV/c)$^2$ (full points),
$Q^{2}$=0.8-1.1~(GeV/c)$^2$ (open squares), and
$Q^{2}$=1.1-1.5~(GeV/c)$^2$ (open triangles). 
Error bars are statistical only, while the bottom band shows the sytematic
error for the lowest $Q^{2}$ bin.
The curves represent our step (A) reference calculations.
Right: $\frac{d\sigma_{v}}{dM_{\pi^{+}\pi^{-}}}$ from CLAS
at $Q^{2}$=0.8-1.1~(GeV/c)$^2$ and $W$=1.7-1.725 GeV
(statistical error bars only). 
The curves represent our step (A) reference calculations,
extrapolated to the edge points.
The dashed line includes all resonances,
the dot-dashed line includes only the non-resonant part, and
the solid line is the full calculation.
}
\label{fig:Xsec_allq2_data_and_nominal_and_pipi}
\end{figure}

Since existing theoretical models \cite{Ose00} are limited
to $W<$1.6 GeV,
we have employed a phenomenological calculation \cite{Mok01}
for a first
   interpretation of the data.  This model describes the reaction 
   $\gamma_{v} p \rightarrow p \pi^{+} \pi^{-} $ in the
kinematic range of interest as a sum of amplitudes for
$\gamma_{v} p \rightarrow \Delta \pi \rightarrow p \pi^{+} \pi^{-} $ and 
$\gamma_{v} p \rightarrow \rho^{0} p \rightarrow p \pi^{+} \pi^{-} $,
% at tree level
while all other possible mechanisms are
parameterized as phase space. 
A detailed treatment was developed 
%\cite{Rip99} 
for the non-resonant
contributions to $\Delta \pi $, while for $\rho p$
production they were described through a diffractive ansatz.
For the resonant part, a total of 12 states, classified as
4$^{*}$ \cite{Pdg96}, with sizeable $\Delta \pi$ and/or $\rho p$ 
decays, were included based on a Breit-Wigner ansatz.
A few model parameters in non-resonant production
were fitted to CLAS data at high $W$, where the non-resonant
part creates a forward
peaking in the angular distributions, and kept fixed 
in the subsequent analysis. The phase between resonant 
and non-resonant $\Delta \pi $ mechanisms 
was fitted to the CLAS data as well.
To simplify the fits, we reduced eqn. (1) 
to three single-differential cross sections,
$\frac{d \sigma}{dM_{p \pi^+}}$,
   $\frac{d \sigma}{dM_{\pi^+ \pi^-}}$, and $\frac{d \sigma}{d \cos\theta_{\pi^-}}$,
   by integrating over the other hadronic variables.  
   These three 1-D distributions were 
   then fitted simultaneously.  
   Here $\frac{d \sigma}{dM_{p \pi^+}}$ and
   $\frac{d \sigma}{d \cos\theta_{\pi^-}}$ are both connected with the
   dominant $\Delta^{++} \pi^-$ production reaction, while
   $\frac{d \sigma}{dM_{\pi^+ \pi^-}}$ is connected with $p \rho^0$
   production.
%For all our fits, we calculated a $\chi^{2}$ per data point
%in each $W$ and $Q^{2}$ bin, based on 26 data points 
%from the 3 mentioned differential cross sections
%(2 edge points in each of the mass distributions were excluded as
%the model did not take into account the kinematic smearing
%in the $M_{\pi^{+}\pi^{-}}$ vs $M_{p \pi^{+}}$ Dalitz plot,
%caused by the $W$ bin width). 
%When fitting the data, we calculated 
%a $\chi^{2}$ per degree of freedom ($\chi^{2}/\nu$) 
%for all $Q^{2}$ values and in the restricted
%$W$ range where we focused our analysis, as discussed in the following
%paragraphs. 
For each $W$ and $Q^2$ bin, a total of 26
   data points from the three single-differential cross sections
   were used in our fits. The two edge points in both the $p \pi^{+}$ and
   $\pi^{+} \pi^{-}$ mass distributions were excluded as the model did not
   take into account the kinematic smearing in the $M_{\pi^{+} \pi^{-}}$ versus
   $M_{p \pi^{+}}$ plot caused by the $W$ bin width.
   %The behavior of our fits was evaluated both on the basis of
%$\chi^{2}$ in each $W$ and $Q^{2}$ bin and by calculating
%the average $\chi^{2}$ over $W$ and $Q^{2}$ in the restricted
%$W$ range where we focused our analysis, as discussed in the following
%paragraphs.

The physics analysis included the following steps:
(A) We produced reference curves using the available 
%PDG
information on the
   $N^*$ and $\Delta$ resonances in 1.2-2 GeV mass range.
%\cite{Pdg96}
Discrepancies between the CLAS data and our calculation were observed,
which led to the subsequent steps B and C.
(B) Data around $W$=1.7 GeV were fitted using the 
   known resonances in the PDG but allowing the resonance parameters 
   to vary in a number of ways.  The best fit, corresponding to a
   prominent $P_{13}$ partial wave, could be
attributed to the PDG $P_{13}$(1720) resonance, but with parameters significantly modified
   from the PDG values.
(C) As an alternative description, we introduced a new baryon state around
1.7~GeV.
In what follows we describe each of the steps above in more detail.

Step (A) - To produce our reference curves,
   the $Q^2$ evolution of the $A_{1/2}$ and $A_{3/2}$ electromagnetic
   couplings for the states was taken either from parameterizations
   of existing data \cite{Bur94}, or from 
Single Quark Transition Model (SQTM) fits \cite{Bur94} where no data were 
available.
For the $P_{11}$(1440) (Roper), given the scarce available data,
the amplitude $A_{1/2}$ was taken from a 
Non-Relativistic Quark Model (NRQM) \cite{Clo90}.
Partial $LS$ decay widths were taken from
a previous analysis of hadronic data \cite{Man92} and renormalized
to the total widths from Ref. \cite{Pdg96}.
Results from step (A) are reported 
in Figs.~\ref{fig:sqtm1} to~\ref{fig:sqtm5} for the specific one-dimensional
differential cross sections analysed and for a sample of $W$ bins.
In Fig.~\ref{fig:Xsec_allq2_data_and_nominal_and_pipi} (left), we report
the total cross section data and the corresponding curves from step (A).
As demonstrated by the plots, our calculation, even without performing
any adjustment of the resonance parameters, is able to give a good accounting
of the main features of the data in a wide $W$ and $Q^2$ region.
The total cross section strength for $W<$ 1.65 GeV 
(except for the region close to threshold), and
for $W>$ 1.8 GeV is well reproduced. 
However, a strong discrepancy is evident at $W$ around 1.7 GeV.
Moreover, at this energy the reference curve
exhibits a strong peak in the $\pi^+ \pi^-$
   invariant mass 
(Fig.~\ref{fig:Xsec_allq2_data_and_nominal_and_pipi}, right), 
connected to sizeable
$\rho$ meson production. This contribution was traced back to the 70-91\% 
branching ratio of the $P_{13}$(1720) into this channel \cite{Pdg96,Man92,Dyt00}. 
%%
%\begin{figure}[h]
%\vspace{5.cm}                                                                  
%\special{psfile=sqtmrb095_171.ps hscale=48 vscale=30 hoffset=-10 voffset=-50}
%\caption[]{Cross section for 
%$\gamma_{v} p \rightarrow p \pi^{+} \pi^{-}$ 
%at $Q^{2}$=0.8-1.1~(GeV/c)$^2$ and $W$=1.7-1.725 GeV, 
%differential in 
%%%%(top to bottom):  
%%%%p$\pi^{+}$ invariant mass; 
%the $\pi^{+}$$\pi^{-}$ invariant mass.
%%%%$\theta_{\pi^{-}}$ CM angle.
%The data from CLAS include statistical error bars only. 
%The curves are our reference calculations described in step (A)
%of the text.
%The dashed line includes all resonances,
%the dot-dashed line includes only the non-resonant part, and
%the solid line is the full calculation.
%}
%\label{fig:Xsec_diff1}
%\end{figure}
%%

Step (B) - We then considered whether the observed
discrepancy around 1.7 GeV could be accomodated by
varying the electromagnetic excitation of one or more
of the PDG states. In our code, interference
effects between continuum and resonances like those reported
in Ref.~\cite{Ose00}, are taken into account
automatically inside the model frame \cite{Mok01}. Therefore, our
investigation at this stage was including the possibility
of accounting for the 1.7 GeV structure via interference effects,
although the peaking of such an interference pattern at the same $W$ for 
all $Q^2$ bins would be rather surprising.
%%%%%%%%%%%%%%%%%%%%%%%%%%%%%%%%%%%%%%%%%%%%%%%%%%%%%%%%%%%%%%%%%%%%%%
%%%%%%%%%Starting from the above mentioned reference values, 
%%%%%%%%%the parameters of various states were varied in order
%%%%%%%%%to fit the CLAS data. In this discussion, we restrict
%%%%%%%%%ourselves to the discrepancy around 1.7 GeV and the few
%%%%%%%%%resonant states relevant in this energy region.
%%%%%%%%%%%%%%%%%%%%%%%%%%%%%%%%%%%%%%%%%%%%%%%%%%%%%%%%%%%%%%%%%%%%%%
%Therefore all $\chi^{2}$ values presented are averaged over the 3 $W$ bins 
%between 1.69 and 1.74 GeV and over the 3 $Q^{2}$ bins (234 data points). 
All fit $\chi^2/\nu$ values were calculated from the 8 $W$ bins 
between 1.64 and 1.81 GeV and from the 3 $Q^{2}$ bins (624 data points). 
The number of free parameters ranged from 11 to 32,
depending on the fit,
corresponding to $\nu$=613 to 592 degrees of freedom.  
%(free photocouplings were counted 3 times for the 3 $Q^{2}$ bins,
%as they represent $Q^{2}$-dependent quantities).
Assuming the resonance properties given by the PDG,
the bump at
   about $W$=1.7 GeV cannot be due to the
$D_{15}$(1675), $F_{15}$(1680), or $D_{33}$(1700) states; 
the first because its well known position cannot match the peak;
the second because of its well known position and 
photocouplings \cite{Bur01};
the third due to its large width ($\sim$300 MeV).
The remaining possibilities from the PDG 
are the $D_{13}$(1700), the $P_{13}$(1720), and the $P_{11}$(1710)
(the latter was not included in step (A)), 
as there are no data available on the $Q^2$ 
dependence of $A_{1/2}$ or $A_{3/2}$~\cite{Bur01}. 
If no configuration mixing 
occurs, the $D_{13}$(1700) cannot be excited in the SQTM
from proton targets,
while the SQTM prediction for the $P_{13}$(1720) 
relies on ad hoc assumptions.
%, being input data from states
%in the same multiplet not sufficient \cite{Bur94}.
According to the literature \cite{Pdg96,Man92,Dyt00}, 
hadronic couplings of the $D_{13}$(1700) and the total width
of the $P_{11}$(1710)
are poorly known, while the $P_{13}$(1720) hadronic
parameters should be better established.
%Therefore, we decided to investigate all these possibilities
%for fitting the bump.
Several other partial waves were investigated
in step (C). 
%Therefore
%we performed two separate fits where one or the other state was producing
%the bump, by varying both the
%electromagnetic and hadronic couplings.
%
\begin{figure}[h]
\vspace{6.0cm}                                                                  
\includegraphics{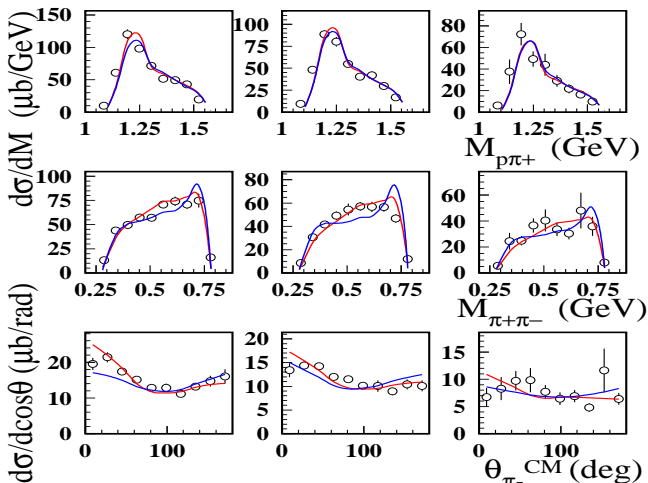}
%blinp13065-130_171_P11.ps hscale=48 vscale=35 hoffset=-12 voffset=-60}
\caption[]{$\frac{d\sigma_{v}}{dM_{p \pi^{+}}}$, 
$\frac{d\sigma_{v}}{dM_{\pi^{+}\pi^{-}}}$, 
and $\frac{d\sigma_{v}}{dcos\theta_{\pi^{-}}}$
from CLAS (from top to bottom) at $W$=1.7-1.725~GeV
and for the three mentioned $Q^{2}$ intervals (left to right).
The error bars include statistical errors only. 
Curves (see text) correspond to 
the fits (B2)
%step (B) when fitting the 1.7 GeV
%bump with $D_{13}$(1700) 
(red), 
and (B4)
%$P_{13}$(1720) 
(blue), 
%or 
%B(3)
%$P_{11}$(1710) 
%(dotted)
and
are extrapolated to the mass distributions edge points.
}
\label{fig:fit_miss}
\end{figure}
\begin{table}
\caption{PDG $P_{13}$(1720) parameters from fit (B) and new state parameters
from fit (C). Errors are statistical.\label{table:tabres_1}}
\vspace{2mm}
%\begin{ruledtabular}                                                            
\begin{tabular}{|c|c|c|c|c|}  \hline
	          &  M (MeV)		 & $\Gamma$ (MeV)     & $\frac{\Gamma_{\pi \Delta}}{\Gamma}$ (\%) & $\frac{\Gamma_{\rho N}}{\Gamma}$ (\%)       \\ \hline
%		  &  (MeV)		 & (MeV)	      & 	 (\%)		       	  	  &	 (\%)		       	    		\\ \hline
% our fit (B)	  &  1725$\pm$20	 & 109$\pm$20$\pm$??  & 49$\pm$16$\pm$??		  	  & 9$^{+10}_{-9}$$\pm$??	 		\\ \hline
%%%PDG $P_{13}$ (B)  &  1725$\pm$20	 & 114$\pm$19$\pm$29  & 63$\pm$12$\pm$17		  	  & 19$\pm$9$\pm$14		 		\\ \hline
%%% PDG \cite{Pdg96}  &  1650-1750		 &     100-200	      &		absent			  	  & 70-85		   			\\ \hline
%%%new $P_{13}$ (C)  &  1720$\pm$20	 & 88$\pm$17$\pm$25   & 41$\pm$13$\pm$20		  	  & 17$\pm$10$\pm$17		 		\\  \hline
PDG $P_{13}$ (B)  &  1725$\pm$20	 & 114$\pm$19	      & 63$\pm$12			  	  & 19$\pm$9			 		\\ \hline
 PDG \cite{Pdg96}  &  1650-1750		 &     100-200	      &		N/A			  	  & 70-85		   			\\ \hline
new $P_{13}$ (C)  &  1720$\pm$20	 & 88$\pm$17	      & 41$\pm$13			  	  & 17$\pm$10			 		\\  \hline
\end{tabular}	
%\end{ruledtabular}                                                              
\end{table}

\begin{figure}[h]
\vspace{5.cm}                                                                  
\includegraphics{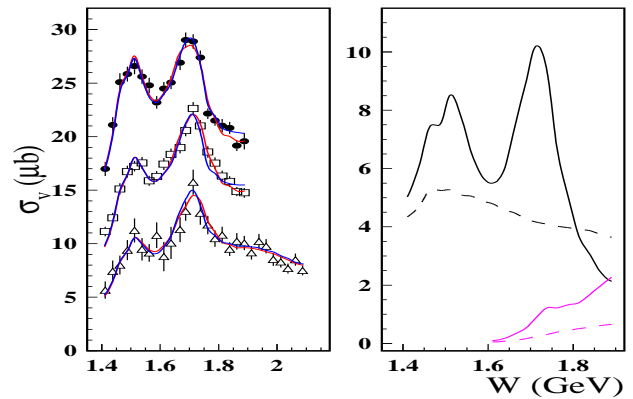}
%fig4_art_new_P11.ps hscale=50 vscale=30 hoffset=-18 voffset=-50}
\caption[]{Left: Total cross section for 
$\gamma_{v} p \rightarrow p \pi^{+} \pi^{-}$ as a function of $W$
from CLAS at the 3 mentioned $Q^{2}$ intervals (see fig.1). 
The error bars are statistical only.
The curves (see text) correspond to 
the fits 
%(B1)
%step (B) when fitting the bump
%with the $D_{13}$(1700) 
%(solid), 
(B2) (red),
%$P_{13}$(1720) 
%(dashed), or 
%(B3)
%$P_{11}$(1710) 
%(dotted). 
%states
and (B4) (blue).
Right: subdivision of the fitted cross section (B2) for
$Q^{2}$=0.5-0.8~(GeV/c)$^2$ into
resonant $\Delta^{++}\pi^{-}$ (black solid), continuum $\Delta^{++}\pi^{-}$
(black dashed), resonant $\rho^{0}p$ (magenta solid),
and continuum $\rho^{0}p$ (magenta dashed). Notice the different
vertical scales.
}
\label{fig:Xsec_allq2_data_and_bestnominal}
\end{figure}

To improve our
   reference curves before fitting the bump at around 1.7 GeV, the
   following steps were carried out:
the $P_{11}$(1440) strength was fitted to our low W data;
the $D_{15}$(1675) and the $D_{13}$(1700) photocouplings (which vanish in the
   SQTM) were replaced by NRQM values from Ref. \cite{Clo90}; 
an empirically established $A_{1/2,3/2}$ SQTM fitting uncertainty or
NRQM uncertainty of 10 or 20\% 
($\sigma$) was applied to all $N^{*}$ states; 
%a 20\% fluctuation was
%applied to the $A_{1/2,3/2}$ NRQM values of the $D_{13}$(1700);
%(such value was empirically established
%based on fitting uncertainties in \cite{Bur94});
the hadronic parameters were allowed to vary for the $D_{13}$(1700) 
according to Ref. \cite{Man92}; and finally, the curves
providing the best $\chi^{2}/\nu$ were selected as the starting points.
First we performed three fits, (B1), (B2), and (B3),
where the photocouplings of only one resonance at a time
were varied.
In (B1), we varied A$_{1/2}$, A$_{3/2}$, hadronic couplings,
and position of the D$_{13}$(1700) in a wide range.
In (B2), the same was done for the P$_{13}$(1720), 
and in (B3) for the P$_{11}$(1710).
In both fits (B2) and (B3),
we also varied the hadronic
parameters and the position of the D$_{13}$(1700)
over a range consistent with their large uncertainties
from Ref. \cite{Man92}. 
%We performed three separate fits, in which the
%hadronic parameters and A$_{/12}$ and A$_{3/2}$ for each
%of the three states, D$_{13}$(1700), P$_{13}$(1720)
%or the P$_{11}$(1710), were separately varied
%over a wide range. In each of these fits, the hadronic
%parameters of the D$_{13}$(1700) were allowed to vary
%over a range consistent with their large uncertainties
%from Ref. \cite{Man92}. 
%While all
%three procedures resulted in reasonable overall fits to
%the W spectrum and Q$^2$
%dependence (Fig.~\ref{fig:Xsec_allq2_data_and_bestnominal}),
%the $\chi^2/\nu$ was 5.2 for (B1),
%the $D_{13}$(1700) 
%4.3 for (B3),
%the $P_{11}$(1710)
%and the descriptions of
%angular distributions and/or the $\pi\pi$ invariant
%mass were poor.
Fits (B1) and (B3) gave a poor description of the data,
with $\chi^2/\nu$=5.2 and 4.3, respectively.
The best fit ($\chi^2/\nu$ =3.4)
was obtained in (B2) (Fig.~\ref{fig:fit_miss}).
%by varying the $P_{13}$(1720). 
However, the resulting values for the branching fractions 
of the P$_{13}$(1720) were significantly
different from previous analyses
reported in the literature and well outside 
the reported errors \cite{Pdg96,Man92,Dyt00}. 
%%%%%%
Starting from (B3), we then performed a final fit, 
(B4), for which the $P_{13}$(1720)
hadronic couplings were fixed from the literature, and  
varying the photocouplings of all three candidate states,
D$_{13}$(1700), P$_{13}$(1720), and $P_{11}$(1710), by 100\% ($\sigma$).
%The new best values of $A_{1/2,3/2}$ coincided 
%with the (B3) results within the errors
No better solution was found, 
the $\chi^2/\nu$ being 4.3 (Fig.~\ref{fig:fit_miss}).  
In Fig.~\ref{fig:Xsec_allq2_data_and_bestnominal} we report the final
comparison of fits (B2) and (B4) with the total cross section data.
Table~\ref{table:tabres_1} shows our results (first row) 
with statistical 
%and 
%systematic 
uncertainties, in comparison with the PDG values (second row). 
%Uncertainties on resonance 
%positions are purely statistical as this parameter is not 
%influenced by systematic errors in the data. 
Our fits were not providing an unambigous
separation of $A_{1/2}$, $A_{3/2}$, 
and the longitudinal $S_{1/2}$, so we report as result
the total photocoupling strength, 
$\sqrt{A_{1/2}^{2} + A_{3/2}^{2}+ S_{1/2}^{2}}$.
The resulting value for the $P_{13}$(1720) fit is reported in the 
first three rows of
Table~\ref{table:tabres_2}.
The errors reflect the statistical uncertainties
   in the data and the correlations among the different resonances.

As discussed above, fitting the data around 1.7 GeV with 
   established baryon states leads either to a poor fit or to a drastic change
in resonance parameters with respect to published results.
In the framework of our analysis, there is
   no way to assess the reliability of the previously determined hadronic 
parameters of the PDG $P_{13}$(1720).
%%%%%%%
The resonant content of the reaction $\pi N \to \pi\pi N$,
                     which is used to obtain the hadronic parameters,
                     may be different from that of reactions initiated by
                     an electromagnetic probe. In particular, the $P_{13}$(1720) 
		     state seen in $\pi N \to \pi\pi N$
   may not be excited in electroproduction, while 
   some other state that decouples from $\pi N$ may be 
   excited electromagnetically.
	This possibility is studied in the next step.
%                     We investigated this pssibility by checking whether
%                     our data could be fitted by including a previously
%                     unreported baryon state, while still accomodating
%                     the known baryons with their published hadronic
%                     properties.
%%%%%%%
%%%%%%%We have to remark at this point that the source of the hadronic
%%%%%%%parameters is the reaction $\pi N \rightarrow \pi \pi N$, while
%%%%%%%the CLAS data are based on an electromagnetic probe in the initial
%%%%%%%channel. Therefore the resonant content
%%%%%%%seen in the two reactions may be different. In particular,
%%%%%%%the conventional $P_{13}$(1720) state may be not well visible
%%%%%%%in two pion electroproduction, while some other state may manifest in
%%%%%%%such reaction as a result of the different entrance channel.
%%%%%%%Therefore we further investigated this possibility,
%%%%%%%checking whether our data could be fitted
%%%%%%%by including another baryon state, not reported
%%%%%%%before, but still accomodating the known baryons with their
%%%%%%%published hadronic properties.
%
\begin{table}
\caption{PDG $P_{13}$(1720) total photocoupling from fit (B2) 
%%%%in comparison with SQTM predictions\cite{Bur94} 
and new state total photocoupling
from fit (C). Errors are statistical.\label{table:tabres_2}}
\vspace{2mm}
%\begin{ruledtabular}                                                            
%%%%%%\begin{tabular}{|c|c|c|c|c|}  \hline
\begin{tabular}{|c|c|c|}  \hline
 step	   & $Q^{2}$	       &    $\sqrt{A_{1/2}^{2} + A_{3/2}^{2}+ S_{1/2}^{2}}$	   \\
	   &(GeV/c)$^{2}$      &($10^{-3}/\sqrt{{\rm GeV}}$)				   \\ \hline
%%%%%% 0.65		   &	 2$\pm$21$\pm$27	  &	     65 		  &   -83$\pm$5$\pm$10	 	 &		-38		       \\ \hline
%%%%%% 0.95		   &	 3$\pm$29$\pm$33	  &	     63 		  &   -63$\pm$8$\pm$11	 	 &		-37		       \\ \hline
%%%%%% 1.30		   &	 2$\pm$12$\pm$13	  &	     61 		  &   -45$\pm$27$\pm$27	 	 &		-35		       \\ \hline
%%%%%% 0.65		   &	 15$\pm$25$\pm$29	  &				  &   -74$\pm$8$\pm$9	 	 &				       \\ \hline
%%%%%% 0.95		   &	 12$\pm$20$\pm$20	  &				  &   -53$\pm$6$\pm$6	 	 &				       \\ \hline
%%%%%% 1.30		   &	 3$\pm$14$\pm$14	  &				  &   -41$\pm$18$\pm$18  	 &				       \\ \hline
 B2		   &	 0.65		   &	 83$\pm$5	     	        	      \\ \hline
 B2		   &	 0.95		   &	 63$\pm$8	     	        	      \\ \hline
 B2		   &	 1.30		   &	 45$\pm$27	     	        	      \\ \hline
 C		   &	 0.65		   &	 76$\pm$9	     	        	      \\ \hline
 C		   &	 0.95		   &	 54$\pm$7	     	        	      \\ \hline
 C		   &	 1.30		   &	 41$\pm$18	     	        	      \\ \hline
\end{tabular}		  
%\end{ruledtabular}                                                              
\end{table}

Step (C) - We investigated whether our
   data could be fitted by including another baryon state, 
while keeping the hadronic parameters of 
the $P_{13}$(1720) as in Refs. \cite{Pdg96,Man92}.
The quantum numbers $S_{I1}$,$P_{I1}$,$P_{I3}$,$D_{I3}$,$D_{I5}$,
$F_{I5}$,$F_{I7}$ were tested on an equal footing, where $I$/2
   is the isospin, 
   undetermined in our measurement.
%Quark model \cite{Cap94} hadronic couplings of the $P_{13}(1870)$ 
%(lowest unobserved resonance) were used as starting point. 
We then simultaneously varied the photocouplings
   and the hadronic parameters of the new state and the $D_{13}$(1700).
   The total decay width of the new state was varied in the range of
   40-600 MeV, while its position was varied from 1.68-1.76 GeV. 
The best fit ($\chi^2/\nu$=3.3) was obtained with a $P_{I3}$ state, 
while keeping
the $P_{13}$(1720) hadronic parameters at published values.
Other partial waves gave a $\chi^{2}/\nu$ $\geq$ 4.2.
Curves obtained from the best fit were nearly identical with the 
%dashed 
red solid lines in
Figs.~\ref{fig:fit_miss} and~\ref{fig:Xsec_allq2_data_and_bestnominal}.
In order to avoid the unobserved $\rho$ production peak
(Fig.~\ref{fig:Xsec_allq2_data_and_nominal_and_pipi}, right), 
the photocouplings of the PDG $P_{13}$(1720)
had to be reduced by about a factor of two with respect to the SQTM prediction,
making its contribution very small.
Instead, in this fit the main contribution to the bump 
came from the new state.
%As a result, we could not definitely exclude an $S_{11}$ state
%($\chi^{2}$=4.9), while $P_{11}$ or spin higher than $\frac{3}{2}$ 
%resulted in a poorer fit ($\chi^{2}> 5$). 
Resonance parameters and total photocoupling value obtained from the assumed new state
are reported in Table~\ref{table:tabres_1} (last row) and~\ref{table:tabres_2} 
(last 3 rows), respectively.

\begin{figure}[h]
\vspace{5.cm}                                                                  
\includegraphics{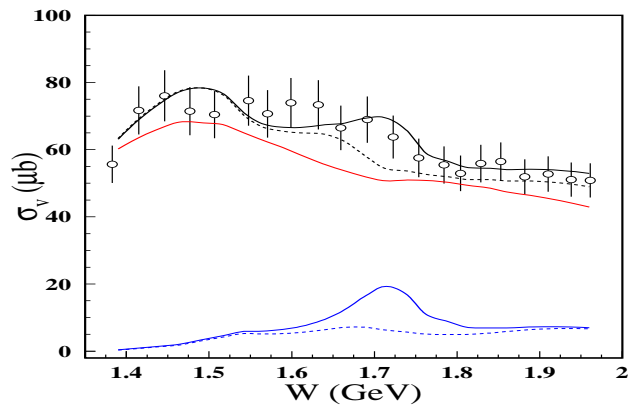}
%fig4_art_new_P11.ps hscale=50 vscale=30 hoffset=-18 voffset=-50}
\caption[]{Total cross section for 
$\gamma p \rightarrow p \pi^{+} \pi^{-}$ as a function of $W$
from Ref.~\cite{ABB68}, together with curves obtained from 
our model. Red solid line: overall contribution from the
non-resonant processes; blue dashed line: resonant
contribution only, without introducing any new state; black dashed line:
full calculation, without new states; 
black solid curve: full calculation, including the new state; blue solid curve:
resonant contribution only, including the new state.
}
\label{fig:Xsec_photonpoint}
\end{figure}

At this point, one could wonder whether a similar discrepancy between
the calculation and the data is present in the real photon case, and
whether introduction of a new state is compatible with the existing 
real photon data as well. To this purpose, we reproduce in Fig.~\ref{fig:Xsec_photonpoint}
the old data from Ref.~\cite{ABB68}, together with curves obtained from 
our model. The red solid curve in the graph represents the overall contribution from the
non-resonant processes, while the blue dashed line corresponds the resonant
contribution only, without introducing any new state. The black dashed line
shows the result of our model when including both non-resonant and resonant
processes, without new states. Finally, the black solid curve shows our fit
of the real photon data when introducing a new state, and the blue solid curve
shows the corresponding modification in the pure resonant excitation curve.
The photocouplings of the new $\frac{3}{2}^+$(1720) state were 
determined extrapolating the values
obtained from the CLAS experiment with virtual
photons. An additional adjustment, within 10 \%, was applied to reproduce the 
invariant mass distributions of the pairs $p \pi^+$ and $\pi^+ \pi^-$, 
measured at $W$=1.7 GeV~\cite{ABB68}.
It is clear how, due to the stronger dominance of the non-resonant mechanisms
at the photon point, as well as due to the particular interference effects,
a new state can be accomodated into the picture. In fact, the agreement between
data and our calculation is even improved by the introduction of the new state.

A second $P_{13}$ state was indeed predicted in Ref. \cite{Cap94}, with a mass 
of 1870 MeV, and in Ref. \cite{Gia01}, with a mass of 1816 MeV. 
The presence of a new three-quark state with the 
same quantum numbers as the conventional
$P_{13}$(1720) in the same mass range would likely lead to strong mixing. 
However, as mentioned above, a different
isospin and/or partial wave cannot be excluded. 
The new state may also have a different internal structure, such as a 
hybrid baryon with excited glue components.
Such a $P_{13}$ hybrid state is predicted in the flux tube model \cite{Pag00}. 
Finally, in Ref.~\cite{Wal03} the existence of 
a $P_{13}$ pentaquark ($qqqq\bar{q}$) configuration with a mass in the
range 1.76-1.78 GeV is predicted.
Yet another possibility is that some resonance parameters established in 
previous analyses may have much larger uncertainties than reported in the 
literature.
In this case, outlined in our step (B),
our analysis would establish new, more precise parameters for a 
known state, and invalidate previous results.

In conclusion, in this paper we presented data on 
the $ep \to e'p \pi^+ \pi^-$ reaction in a wide kinematic range,
with higher quality than any previous double 
pion production experiment.
Our phenomenological calculations using existing PDG parameters 
provided a poor agreement with the new data at $W \sim$ 1700 MeV.  
We explored two alternative interpretations of 
the data. If we dismiss previously established hadronic 
parameters for the $P_{13}$(1720) we can fit the data with
a state having the same spin/parity/isospin but strongly 
different hadronic couplings from the PDG state. 
If, alternatively, we introduce a new state in addition 
to the PDG state with about the same mass, spin $\frac{3}{2}$, and 
positive parity, a good fit is obtained for a state having a 
rather narrow width, a strong $\Delta \pi$ coupling, and a 
small $\rho N$ coupling, while keeping
the PDG $P_{13}$(1720) hadronic parameters at published values. 
In either case we determined the total photocoupling 
%for the PDG $P_{13}$(1720) state 
at $Q^2 > 0$.
A simultaneous analysis of single and 
double-pion processes provides more constraints and may  
help discriminate better between alternative interpretations of 
the observed resonance structure in the CLAS data. Such 
an effort is currently underway.

We would like to acknowledge the outstanding efforts of the 
staff of the Accelerator
and the Physics Divisions at JLab that made this experiment possible. 
This work was supported in part by the Istituto Nazionale di Fisica Nucleare, 
the U.S. Department of Energy and National Science Foundation, 
the French Commissariat \`a l'Energie Atomique,  
and the Korea Science and Engineering Foundation.
U. Thoma acknowledges an ``Emmy Noether'' grant from the 
Deutsche Forschungsgemeinschaft.
The Southeastern Universities Research Association (SURA) operates the
Thomas Jefferson National Accelerator Facility for the United States
Department of Energy under contract DE-AC05-84ER40150.

\end{document}